\documentclass{article}
\usepackage{arxiv}

\usepackage[utf8]{inputenc} 
\usepackage[T1]{fontenc}    
\usepackage{hyperref}       
\usepackage{url}            
\usepackage{booktabs}       
\usepackage{amsfonts}       
\usepackage{nicefrac}       
\usepackage{microtype}      
\usepackage{lipsum}         
\usepackage{graphicx}
\usepackage[numbers]{natbib}
\usepackage{doi}
\usepackage{graphicx}
\usepackage{hyperref}
\usepackage{booktabs,makecell} 

\usepackage{booktabs}
\usepackage[utf8]{inputenc}
\usepackage{amsmath}
\usepackage{graphicx}
\usepackage{wrapfig}
\usepackage{float}
\usepackage{arydshln}
\usepackage{xspace}
\usepackage{algorithm}
\usepackage{algpseudocode}
\usepackage{amsmath,amssymb,amsfonts}
\usepackage{graphicx}
\usepackage[labelformat = parens,justification=centering]{subfig}
\usepackage{textcomp}
\usepackage{xcolor}
\usepackage[normalem]{ulem}
\usepackage{todonotes}
\usepackage{multirow}
\usepackage[gen]{eurosym}
\usepackage{placeins}
\usepackage{cleveref}

\newcommand{\csr}{\textsc{CSR}\xspace}
\renewcommand{\ell}{\textsc{ELL}\xspace}

\newcommand{\coo}{\textsc{COO}\xspace}

\newcommand{\gko}{\textsc{Ginkgo}\xspace}
\newcommand{\gkos}{\textsc{Ginkgo's}\xspace}
\newcommand{\gtwelve}{\textsc{Gen12}\xspace}
\newcommand{\gnine}{\textsc{Gen9}\xspace}

\renewcommand{\gko}{\textsc{Ginkgo}\xspace}
\renewcommand{\gkos}{\textsc{Ginkgo's}\xspace}

\newcommand{\spmv}{\textsc{SpMV}\xspace}

\usepackage{color}
\usepackage[procnames]{listings}
\newcommand{\prettysmall}{\fontsize{6.7}{6.7}\selectfont}
\definecolor{gray98}{rgb}{0.98,0.98,0.98}
\definecolor{gray20}{rgb}{0.20,0.20,0.20}
\definecolor{gray25}{rgb}{0.25,0.25,0.25}
\definecolor{gray16}{rgb}{0.161,0.161,0.161}
\definecolor{gray60}{rgb}{0.6,0.6,0.6}
\definecolor{gray30}{rgb}{0.3,0.3,0.3}
\definecolor{bgray}{RGB}{248, 248, 248}
\definecolor{amgreen}{RGB}{77, 175, 74}
 \definecolor{amblu}{RGB}{72, 88, 102}
\definecolor{amblu}{RGB}{55, 126, 184}
\definecolor{amred}{RGB}{228,26,28}
\begin{document}
\title{Porting a sparse linear algebra math library to Intel GPUs}
\author{
\href{https://orcid.org/0000-0001-5229-3739}{\includegraphics[scale=0.06]{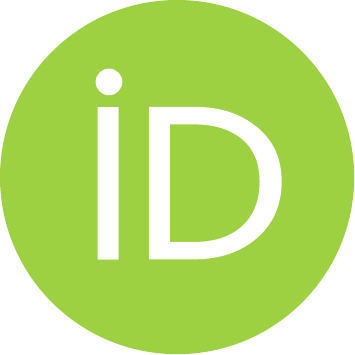}\hspace{1mm}Yuhsiang M. Tsai}\\
Karlsruhe Institute of Technology\\
76131 Karlsruhe, Germany\\
\texttt{yu-hsiang.tsai@kit.edu}\\
\And
\href{https://orcid.org/0000-0002-1560-921X}{\includegraphics[scale=0.06]{orcid.pdf}\hspace{1mm}Terry Cojean}\\
Karlsruhe Institute of Technology\\
76131 Karlsruhe, Germany\\
\texttt{terry.cojean@kit.edu}\\
\And
\href{https://orcid.org/0000-0003-2177-952X}{\includegraphics[scale=0.06]{orcid.pdf}\hspace{1mm}Hartwig Anzt}\\
Karlsruhe Institute of Technology\\
76131 Karlsruhe, Germany\\
University of Tennessee\\
37996 TN, USA\\
\texttt{hanzt@icl.utk.edu}\\
}
%
%
%
%
%


%
%

\hypersetup{
pdftitle={Sparse Linear Algebra on Intel GPUs},
pdfauthor={Yuhsiang M. Tsai, Terry Cojean, Hartwig Anzt},
pdfkeywords={Sparse Matrix Vector Product, Krylov Solvers, GPUs, Intel, oneAPI, DPC++},
}

\maketitle              

\begin{abstract}
With the announcement that the Aurora Supercomputer will be composed of general purpose Intel CPUs complemented by discrete high performance Intel GPUs, and the deployment of the oneAPI ecosystem, Intel has committed to enter the arena of discrete high performance GPUs. A central requirement for the scientific computing community is the availability of production-ready software stacks and a glimpse of the performance they can expect to see on Intel high performance GPUs. In this paper, we present the first platform-portable open source math library supporting Intel GPUs via the DPC++ programming environment. We also benchmark some of the developed sparse linear algebra functionality on different Intel GPUs to assess the efficiency of the DPC++ programming ecosystem to translate raw performance into application performance. Aside from quantifying the efficiency within the hardware-specific roofline model, we also compare against routines providing the same functionality that ship with Intel's oneMKL vendor library.
\keywords{Sparse Matrix Vector Product \and Krylov Solvers \and GPUs \and Intel \and oneAPI \and DPC++}
\end{abstract}
\section{Introduction}
\label{s1-intro}
For a long time, Intel GPUs were almost exclusively available as integrated component of Intel CPU architectures. However, at latest with the announcement that the Aurora Supercomputer will be composed of general purpose Intel CPUs complemented by discrete Intel GPUs and the deployment of the oneAPI ecosystem in cooperation with CodePlay, Intel has committed to enter the arena of discrete high performance GPUs. Other than integrated GPUs, discrete GPUs are usually not exclusively intended to accelerate graphics, but they are designed to also deliver computational power that can be used, e.g., for scientific computations. On the software side, the oneAPI ecosystem promoted by Intel intends to provide a platform for C++ developers to develop code in the DPC++ language that can be executed on any Intel device, including CPUs, GPUs, and FPGAs.

In 2020, Intel released the Intel generation 12 Intel\textsuperscript{\tiny\textregistered} Iris\textsuperscript{\tiny\textregistered} Xe Graphics GPU codename {\emph{DG1}}, an architecture more powerful than the Intel generation 9 integrated GPU deployed in many systems, and with full support of the oneAPI ecosystem. As this GPU may be spearheading the development of Intel's discrete GPU line, we assess the performance this GPU can achieve in numerical calculations. Specifically, we develop a DPC++ backend for the \gko open source math library, and benchmark the developed functionality on different Intel GPU architectures. As \gkos main focus is on sparse linear algebra, we assess the performance of the sparse matrix vector product (\spmv) and iterative Krylov solvers within the hardware-specific performance limits imposed by arithmetic peak performance and memory bandwidth. We consider both double precision and single precision computations and compare against Intel's vendor library oneMKL designed for the oneAPI ecosystem.

Up to our knowledge, we are the first to present the functionality and performance of an open source math library on Intel discrete GPUs. We structure the paper into the following sections:
In \Cref{sec:gko}, we introduce the \gko open source library and its design for platform portability. In \Cref{sec:dpcpp}, we introduce the oneAPI ecosystem and the DPC++ programming environment. In \Cref{sec:porting}, we discuss some aspects of adding a DPC++ backend to \gko for portability to Intel GPUs. For convenience, we briefly recall in \Cref{sec:numerics} the functionality and some key aspects of the algorithms we utilize in our experimental evaluation. This performance evaluation is presented in \Cref{sec:experiments}: we initially benchmark the both the Intel generation 9 and 12 GPUs in terms of feasible bandwidth and peak performance to derive a roofline model, then evaluate the performance of \gkos \spmv kernels (also in comparison to the \spmv routine available in the oneMKL vendor library), and finally assess the performance of \gkos Krylov solvers. For completeness, we include performance results using \gkos other backends on high-end AMD and NVIDIA hardware to demonstrate the (performance) portability of the \gko library. We conclude with a summary of the porting and performance experiences on the first discrete Intel GPU in \Cref{sec:conclusions}.

\section{\gko design}
\label{sec:gko}
\gko is a GPU-focused cross-platform linear operator library focusing on sparse linear algebra~\citep{ginkgoarxiv,ginkgojoss}. The library design is
guided by combining ecosystem extensibility with heavy,
architecture-specific kernel optimization using the platform-native
languages CUDA (NVIDIA GPUs), HIP (AMD GPUs), or OpenMP (Intel/AMD/ARM
multicore)~\cite{cojean2020ginkgo}. The software development cycle ensures production-quality code by featuring unit testing, automated configuration and installation,
Doxygen code documentation, as well as a continuous integration and
continuous benchmarking framework~\cite{gpe}.
\gko provides a comprehensive set of sparse BLAS operations, iterative solvers including many Krylov methods, standard and advanced preconditioning techniques, and cutting-edge mixed precision methods~\cite{toms-adaptiveblockjacobi}.

\begin{figure}[!b]
  \centering
  \includegraphics[width=0.9\linewidth]{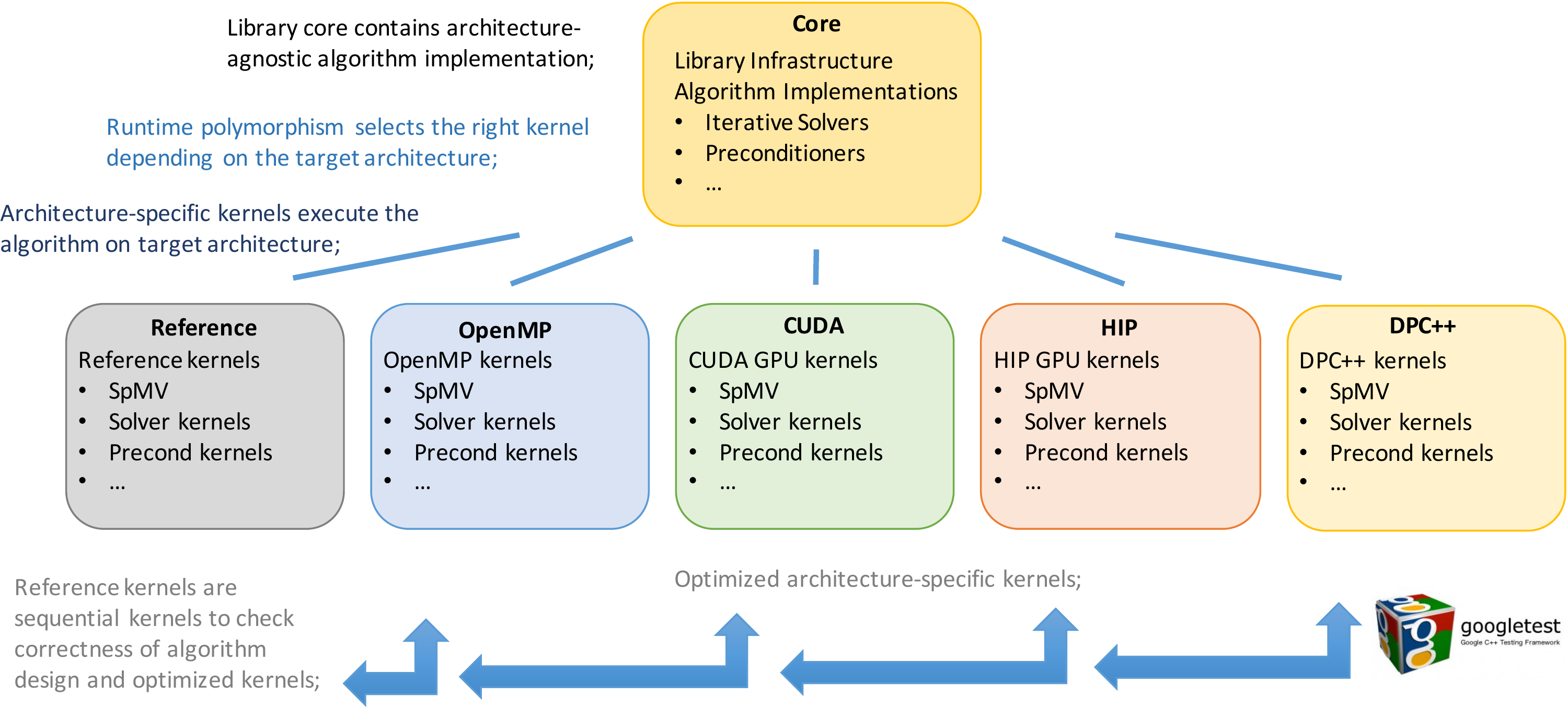}
  \caption{The \gko library design overview.}
  \label{fig:gko_desgin}
\end{figure}

A high-level overview of \gko{}'s software architecture is visualized in 
\Cref{fig:gko_desgin}. The library design collects all classes and generic 
algorithm skeletons in the ``core'' library which, however, is useless without 
the driver kernels available in the ``omp'', ``cuda'', ``hip'', and ``reference'' 
backends. We note that ``reference'' contains sequential CPU kernels used to 
validate the correctness of the algorithms and as reference implementation for 
the unit tests realized using the googletest \cite{googletest} framework.
We note that the ``cuda'' and ``hip'' backends are very similar in kernel design, and we therefore have ``shared'' kernels that are identical for the NVIDIA and AMD GPUs up to kernel configuration parameters~\cite{Tsai2020PreparingGF}.
Extending 
\gko{}'s scope to support Intel GPUs via the DPC++ language, we add the ``dpcpp'' backend containing the kernels in the DPC++ language.

To reduce the effort of adding a DPC++ backend, we use the same 
base components of \gko{} like \texttt{config}, \texttt{binding},  
\texttt{executor}, \texttt{types} and \texttt{operations},
which we only extend and adapt to support DPC++.
\begin{itemize}
\item \texttt{config}: hardware-specific information like warp size, 
lane\_mask\_type, etc.;
\item \texttt{binding}: the C++ style overloaded interface to vendors' BLAS and sparse BLAS
library and the exception calls of the kernels not implemented;
\item \texttt{executor}: the ``handle'' controlling the kernel execution, all form of
interactions with the hardware such as memory allocations and the ability to switch
the execution space (hardware backend);
\item \texttt{types}: the type of kernel variables and the conversion between library 
variables and kernel variables;
\item \texttt{operations}: a class aggregating all the possible kernel implementations
  such as reference, omp, cuda, hip, and dpc++, which allows to switch between
  implementations at runtime when changing the executor type used.
\end{itemize}

\section{The oneAPI Programming Ecosystem}
\label{sec:dpcpp}

oneAPI\footnote{\url{https://spec.oneApi.com/versions/latest/index.html}} is an open and free programming ecosystem which aims at providing portability across a wide range of hardware platforms from different architecture generations and vendors. The oneAPI software stack is structured with the new DPC++ programming language at its core, accompanied by several libraries to ease parallel application programming.

DPC++ is a community-driven (open-source) language based on on the ISO C++ and Khronos' SYCL standards. The concept of DPC++ is to enhance the SYCL~\cite{sycl} ecosystem with several additions that aim at improving the performance on modern hardware, improving usability, and simplifying the porting of classical CUDA code to the DPC++ language. Compared to SYCL, two relevant features of the DPC++ ecosystem are\footnote{Both extensions are now also part of the SYCL 2020 Provisional Specification: \url{https://www.khronos.org/news/press/khronos-releases-sycl-2020-provisional-specification}}: 1) DPC++ introduces a new subgroup concept which can be used inside kernels. This concept is equivalent to CUDA subwarps (or SIMD on CPUs) and allows optimized routines such as subgroup based shuffles. In the \gko library, we make extensive use of this capability to boost the performance. 2) DPC++ adds a new Unified Shared Memory (USM) model which provides new \texttt{malloc\_host} and \texttt{malloc\_device} operations to allocate memory which can either be accessed both by host or device, or respectively accessed by a device only. Additionally, the new SYCL \texttt{queue} extensions
facilitate the porting of CUDA code as well as memory control. Indeed, in pure SYCL, memory copies are entirely asynchronous and hidden from the user, since the SYCL programming model is based on tasking with automatic discovery of task dependencies.

Another important aspect of oneAPI and DPC++ is that they adopt platform portability as the central design concept. Already the fact that DPC++ is based on SYCL (which leverages the OpenCL's runtime and SPIRV's intermediate kernel representation) provides portability to a variety of hardware. On top of this, DPC++ develops a plugin API which allows to develop new backends and switch dynamically between them\footnote{\url{https://intel.github.io/llvm-docs/PluginInterface.html}}.
Currently, DPC++ supports the standard OpenCL backend, a new Level Zero backend which is the backend of choice for Intel hardware\footnote{\url{https://spec.oneApi.com/level-zero/latest/core/INTRO.html}}, and an experimental CUDA backend for targeting CUDA-enabled GPUs. As our goal is to provide high performance sparse linear algebra functionality on Intel GPUs, we focus on the Intel Level Zero backend of DPC++.

\begin{figure}[!t]
\begin{center}
\begin{lstlisting}[language=C,caption={Small example of a SYCL/DPC++ code with a dummy kernel.},label={code:dpcppsample},morekeywords={submit, memcpy, malloc_device}]
#include <CL/sycl.hpp>

int main()
{
  std::vector<int> A(global_range);
  std::vector<int> B(1);

  {
    // Create sycl buffer. Destructed at the end of the scope
    sycl::buffer<sycl::cl_int, 1> bufA(A.data(), A.size());
    sycl::buffer<sycl::cl_int, 1> bufB(B.data(), B.size());

    // Select the default hardware, could be GPU, CPU, ...
    sycl::queue myQueue{};

    // Control the queue submission through a handler
    myQueue.submit([&](sycl::handler &cgh) {
      // Declare buffer access policies for this kernel
      auto accA = bufA.get_access<sycl::access::mode::discard_write>(cgh);
      auto accB = bufB.get_access<sycl::access::mode::atomic>(cgh);
      // Actual kernel submission
      cgh.parallel_for<class hello_world>(
        sycl::range<1>(global_range),
          [=](sycl::id<1> idx) {
            accA[idx] = idx[0]*2;
            accB[0].fetch_add(idx[0]);
      }); // End of the kernel function
    }); // End of the queue commands
  } // End of scope. Synchronizes the previous queue.

  // Now that the data is synchronized, print the buffer
  for (size_t i = 0; i < global_range; i++)
    std::cout << "A[ " << i << " ] = " << A[i] << std::endl;
  std::cout << "The sum of all ranks is: " << B[0] << std::endl;
}
\end{lstlisting}
\end{center}
\label{fig:dpcppsample}
\end{figure}

In Listing~\ref{code:dpcppsample}, we show a minimal example of a SYCL/DPC++ code in a classical use case. In line 10-11, previously declared data is wrapped into a \texttt{sycl::buffer} to enable automatic memory management. In this example, the \texttt{sycl::queue} declared in line 14 automatically selects the execution hardware. In general, the hardware selection can also be controlled explicitly. In line 17-28, the submission of a kernel is controlled through a command group handler. This allows to define accessors for the data in lines 19 and 20. These accessors declare the data access policy of the previous buffers and allow the runtime to automatically infer which data transfers (host/device) are required. Lines 22-27 contain the actual kernel declaration. The accessors are used to write to the previous buffers. 
Taking the C++ principles, at the end of the kernel, DPC++ automatically transfers the buffers back to the vectors A, B, destroys the buffers and synchronizes the queue. As a result, after kernel completion, the (modified) vectors A and B can again be accessed transparently, see lines 31-34.
\section{Porting to the DPC++ ecosystem}
\label{sec:porting}
Porting CUDA code to the DPC++ ecosystem requires to acknowledge that the SYCL-based DPC++ ecosystem is expressing algorithms in terms of tasks and their dependencies, which requires a fundamentally-different code structure. For the porting process, Intel provides the ``DPC++ Compatibility Tool'' (DPCT) that is designed to migrate CUDA code into compilable DPC++ code. DPCT is not expected to automatically generate a DPC++ ``production-ready'' executable code, but ``ready-to-compilation'' and it requires the developer's attention and effort in fixing converting issues and tuning it to reach performance goals. However, with oneAPI still being in its early stages, DPCT still has some flaws and failures, and we develop a customized porting workflow using the DPC++ Compatibility Tool at its core, but embedding it into a framework that weakens some DPCT prerequisites and prevents incorrect code conversion. In general, DPCT requires not only knowledge of the functionality of a to-be-converted kernel, but also knowledge of the complete library and its design. This requirement is hard to fulfill in practice, as for complex libraries, the dependency analysis may exceed the DPCT capabilities. Additionally, many libraries do not aim at converting all code to DPC++, but only a subset to enable the dedicated execution of specific kernels on DPC++-enabled accelerators. In \Cref{sec:isolated_modification}, we demonstrate how we isolate kernels to be converted by DPCT from the rest of the library. Another flaw of the early version of the DPCT is that it typically fails to convert CUDA code making use of atomic operations or the cooperative group functionality. As \gko implementations aim at executing close to the hardware-induced limits, we make heavy use of atomic- and cooperative group operations. In \Cref{sec:dpct_workaround} we demonstrate how we prevent DPCT from executing an incorrect conversion of these operations such that we can convert them using a customized script. To simplify the maintenance of the platform-portable \gko library, our customized porting workflow also uses some abstraction to make the DPC++ code in this first version look more similar to CUDA/ HIP code. We note that this design choice is reflecting that the developers of \gko are currently used to designing GPU kernels in CUDA, but it may not be preferred by developers used to programming in task-based languages. We elaborate on how we preserve much of the CUDA/ HIP code style in \Cref{sec:cuda_familiar}.

\subsection{Isolated Modification}
\label{sec:isolated_modification}
Unfortunately, DPCT needs to know the definition of all functions related to the target file. Otherwise, when running into a function without definition in the target file, DPCT returns an error message. Furthermore, DPCT by default converts all files related to the target file containing any CUDA code that are located in the same folder like the target file\footnote{Also files in subfolders are automatically converted by DPCT if they contain CUDA code.}.
To prevent DPCT from converting files that we do not want to be converted, we have to artificially restrict the conversion to the target files. We achieve this by copying the target files into a temporary folder and considering the rest of the \gko software as a system library. After the successful conversion of the target file, we copy the file back to the correct destination in the new DPC++ submodule.

By isolating the target files, we indeed avoid additional changes and unexpected errors, but we also lose the DPCT ability to transform CUDA kernel indexing into the DPC++ \texttt{nd\_item\textless3\textgreater} equivalent. As a workaround, we copy simple headers to the working directory containing the thread\_id computation helper functions of the CUDA code such that DPCT can recognize them and transform them into the DPC++ equivalent. 
Unfortunately, this workaround works well only if DPCT converts all code correctly. If DPCT fails to convert some files or function definitions live outside the target files, we need to add a fake interface. Examples where the DPCT conversion does not meet our requirements are our custom DPC++ cooperative group interface and the DPC++ CUDA-like dim3 interface which allows to use CUDA-like block and grid kernel instantiation instead of the DPC++ \texttt{nd\_range}. For those, we prevent DPCT from applying any conversion steps but keep DPCT's functionality to add the \texttt{nd\_item\textless3\textgreater} launch parameters.

\subsection{Workaround for Atomic Operations and Cooperative Groups}
\label{sec:dpct_workaround}
DPC++ provides a subgroup interface featuring shuffle operations. However, this interface is different from CUDA's cooperative group design as it requires the subgroup size as a function attribute and does not allow for different subgroup sizes in the same global group.
Based on the DPC++ subgroup interface, we implement our own DPC++ cooperative group interface. 
Specifically, to remove the need for an additional function attribute, we add the \texttt{item\_ct1} function argument into the group constructor. As the remaining function arguments are identical to the CUDA cooperative group function arguments, we therewith achieve a high level of interface similarity. 

\begin{figure}
    \centering
    \includegraphics[width=\textwidth]{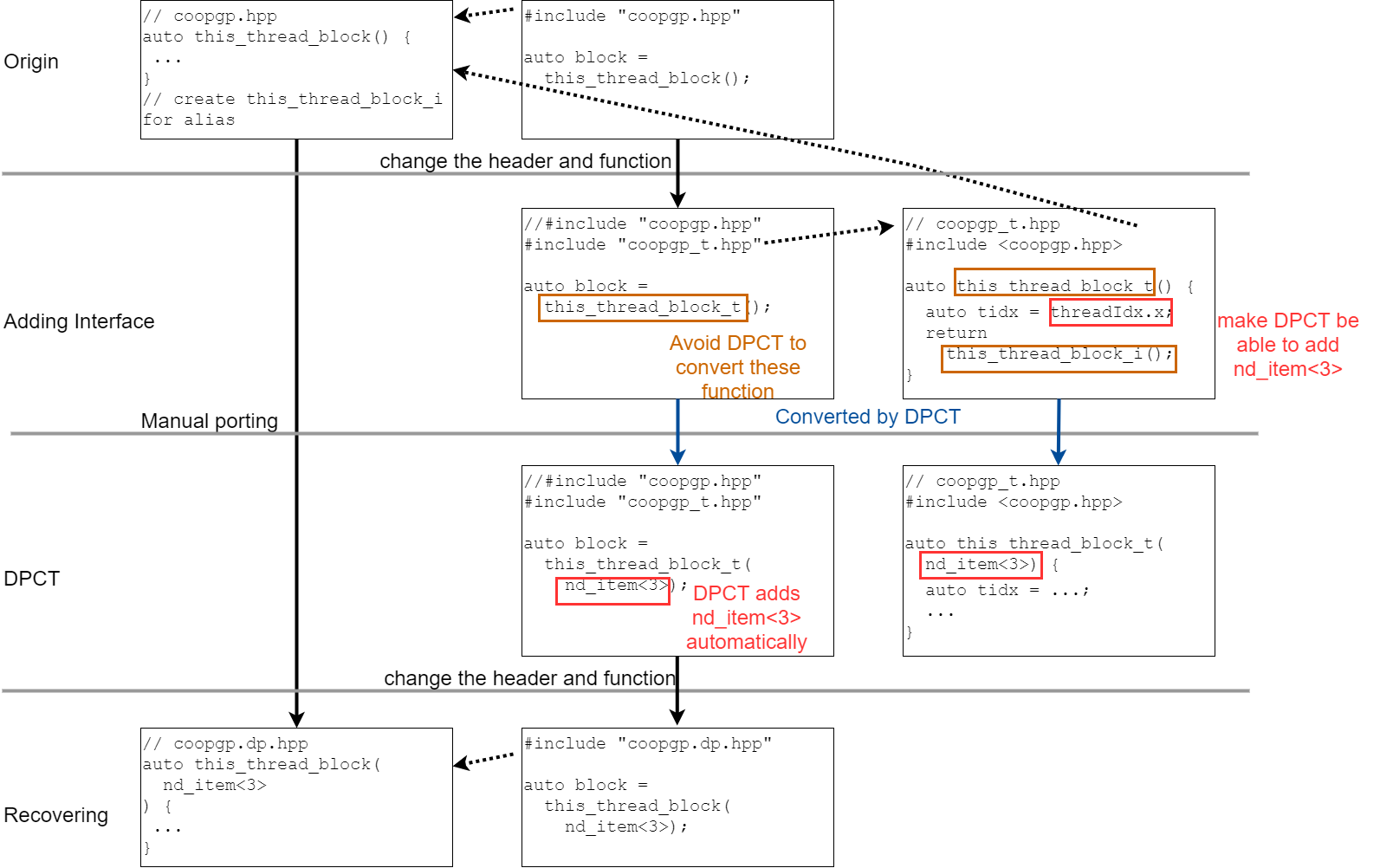}
    \caption{Summary of the workflow used to port the cooperative groups functionality and isolating effort such that we get the correct converted DPC++ codes.}
    \label{fig:dpct_workaround_with_coop}
\end{figure}

A notable difference to CUDA is that DPC++ does not support subgroup vote functions like ``ballot'', ``any'', or other group operations yet. To emulate this functionality, we need to use a subgroup reduction or some algorithms provided by the oneAPI groups to emulate these vote functions in a subgroup setting. This lack of native support may affect the performance of kernels relying on these subgroup operations.
We visualize the workflow we use to port code making use of the cooperative group functionality in \Cref{fig:dpct_workaround_with_coop}. This workflow composes four steps:
\begin{enumerate}
    \item Origin: We need to prepare an alias to the cooperative group function such that DPCT does not catch the keyword. We create this alias in a fake cooperative group header we only use during the porting process.
    \item Adding Interface: As explained in \Cref{sec:isolated_modification}, we need to isolate the files to prevent DPCT from changing other files. During this process we add the simple interface including \texttt{threadIdx.x} and make use of the alias function. Note that for the conversion to succeed, it is required to return the same type as the original CUDA type, which we need to extract from the CUDA cooperative group function \texttt{this\_thread\_block}.
    \item DPCT: Apply DPCT on the previously prepared files. As we add the \texttt{threadIdx.x} indexing to the function, DPCT will automatically generate the \texttt{nd\_item\textless3\textgreater} indexing for us.
    \item Recovering: During this step, we change the related cooperative group functions and headers to the actual DPC++ equivalent. We implement a complete header file which ports all the cooperative group functionality to DPC++.
\end{enumerate}

\begin{figure}
    \centering
    \subfloat[CUDA small cooperative group example\label{fig:cuda_coop}
    ]{\includegraphics[width=.45\columnwidth]{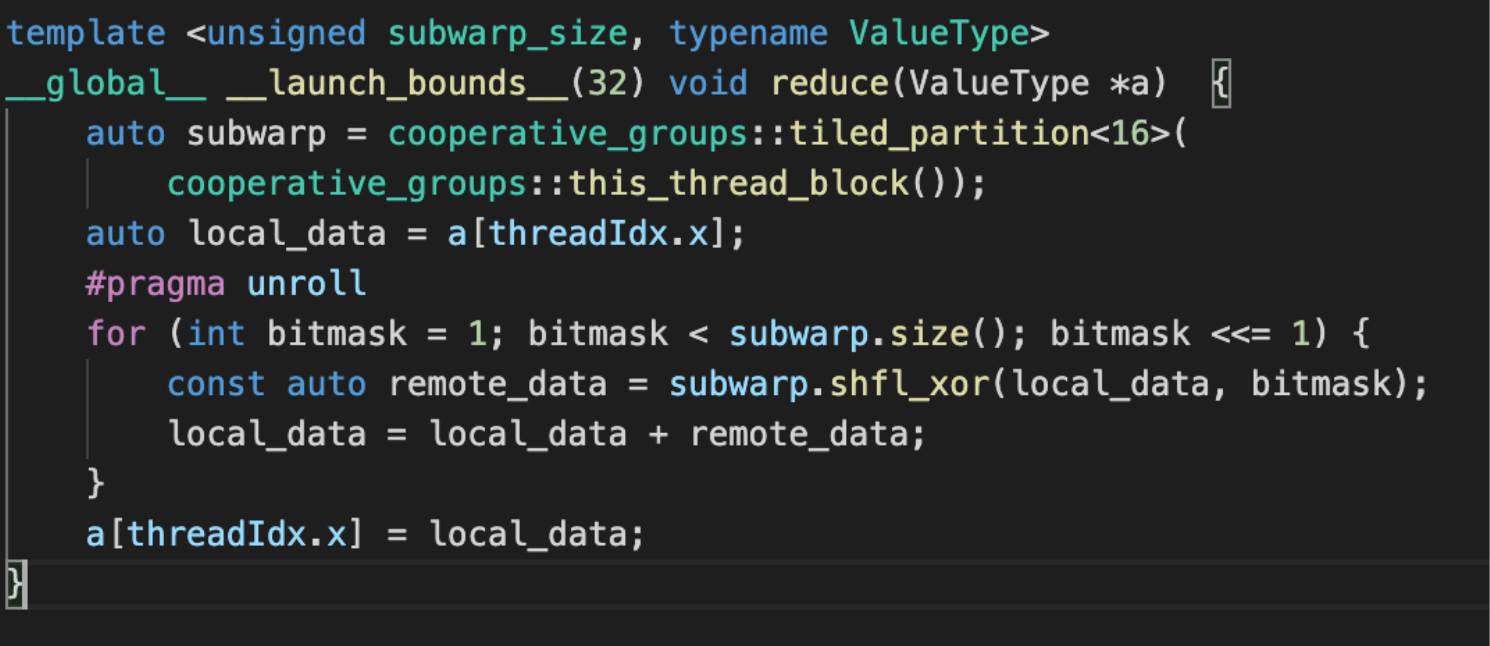}
    }
    \subfloat[DPCT conversion fails and reports error\label{fig:dpct_fail_coop}
    ]{\includegraphics[width=.45\columnwidth]{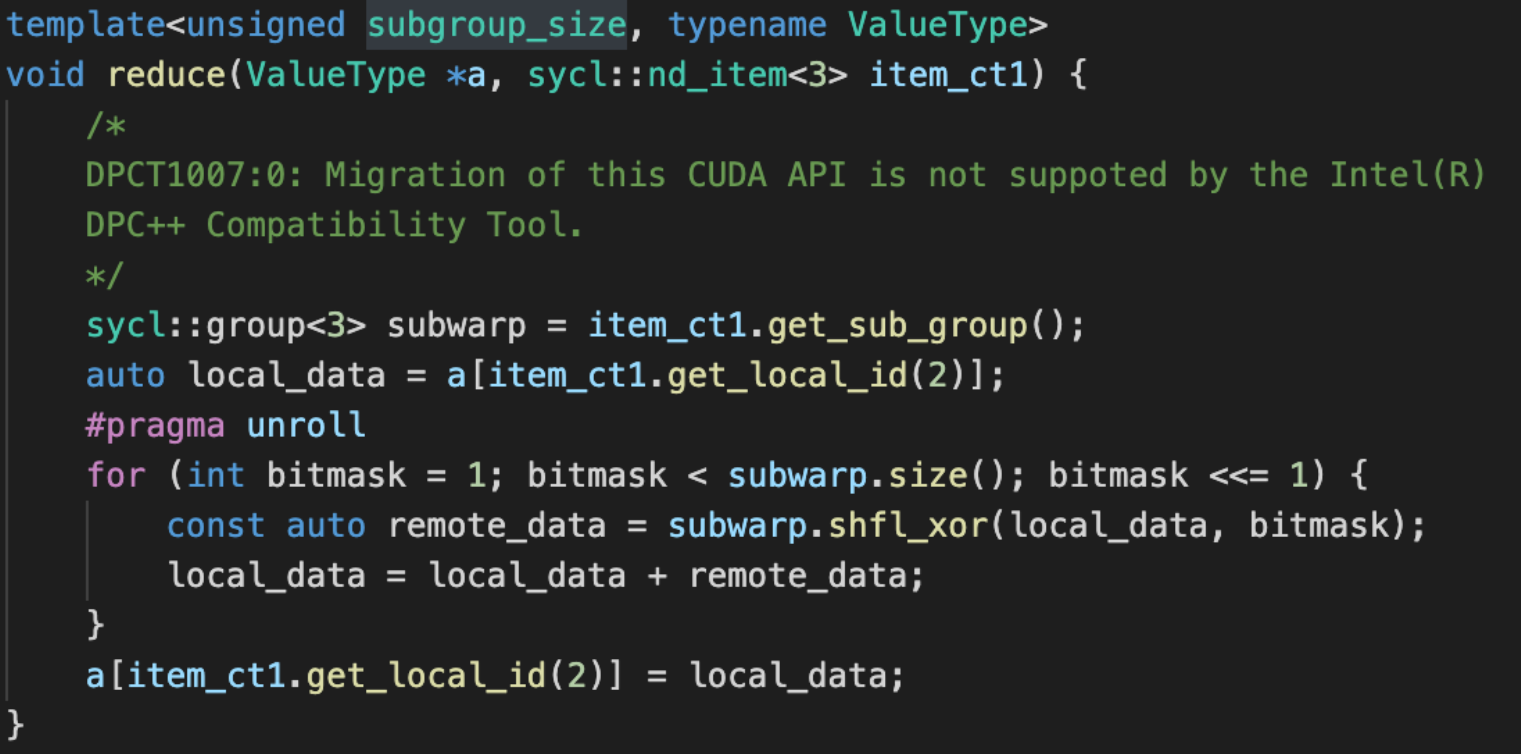}
    }\\
    \subfloat[Manual DPC++ subgroup implementation. The orange boxes are the main difference from CUDA\label{fig:dpcpp_coop}
    ]{\includegraphics[width=.45\columnwidth]{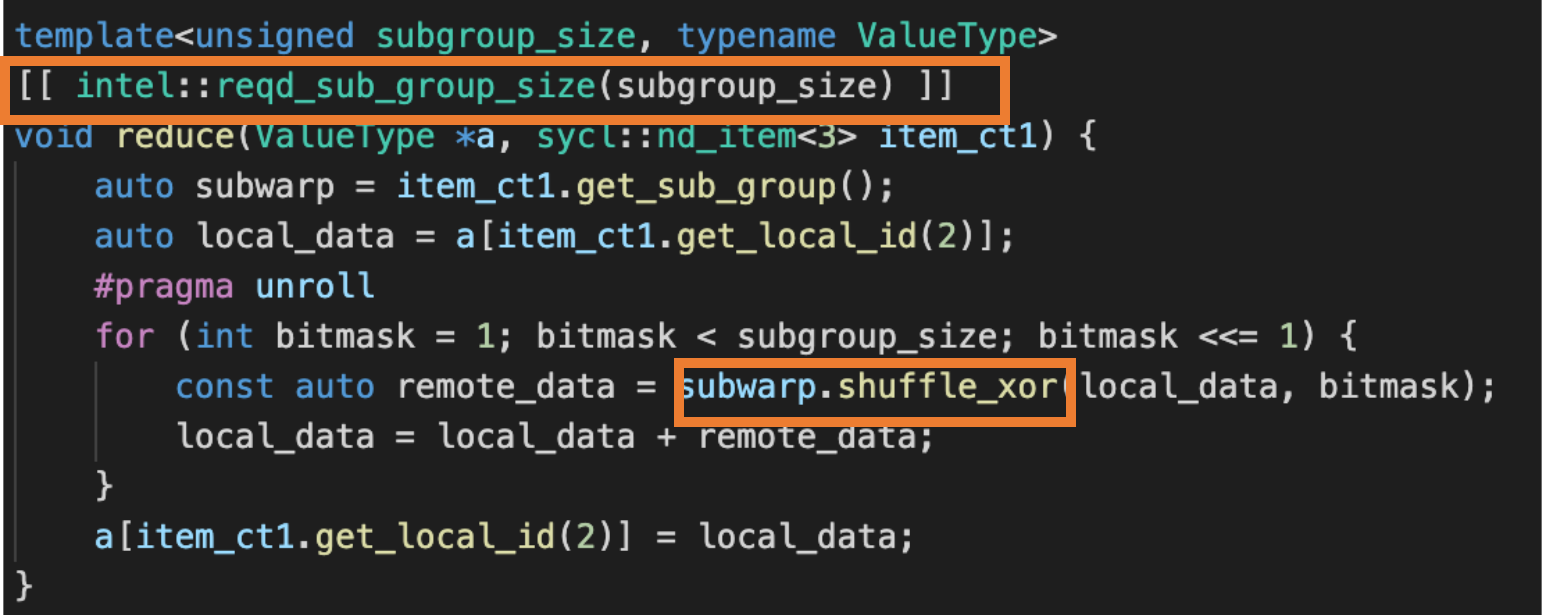}
    }
    \subfloat[The result converted by our porting script\label{fig:gko_ported_coop}
    ]{\includegraphics[width=.45\columnwidth]{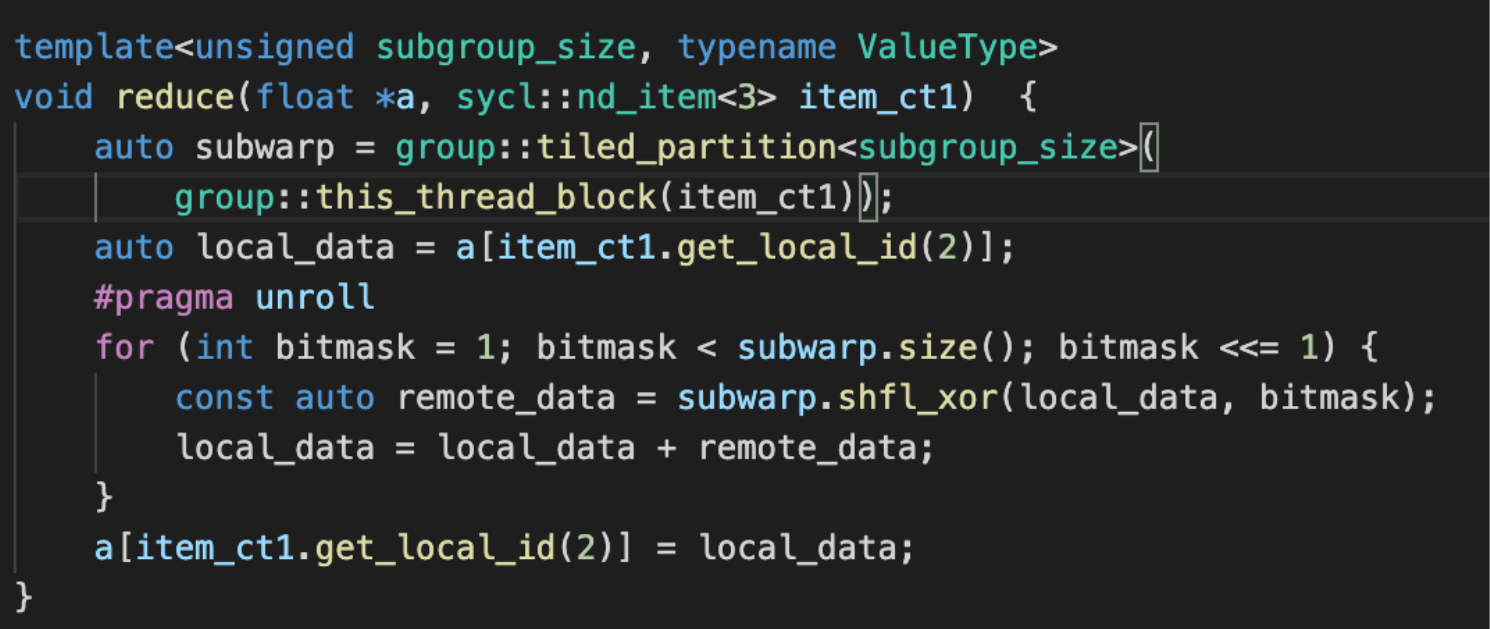}
    }
    \caption{The cooperative group example}
    \label{fig:cooperative_group}
\end{figure}

We show in \Cref{fig:cooperative_group} the final result of the porting workflow on a toy example featuring the use of cooperative groups. For the small example code in \Cref{fig:cuda_coop}, if we do not isolate the code, DPCT will throw an error like \Cref{fig:dpct_fail_coop} once encountering the cooperative group keyword. A manual implementation of the cooperative group equivalent kernel is shown in \Cref{fig:dpcpp_coop}. Our porting workflow generates the code shown in \Cref{fig:gko_ported_coop}, which is almost identical to the original CUDA code \Cref{fig:cuda_coop}.

The conversion of CUDA atomics to DPC++ atomics is challenging as the conversion needs to recognize the data location and decide whether the DPC++ atomics operate on local or global memory. DPCT generally succeeds in this automated memory detection, however, there are two aspects that require us to create a workaround: 1) at the time of writing, DPCT fails to correctly convert atomic operations on local memory\footnote{We reported this issue to the Intel Development Team and they are working on a bugfix.}; and 2) DPC++ does not provide atomics for complex floating point numbers\footnote{Atomic operations on complex numbers is correct only if a single modification is applied at a time.}. We prevent DPCT from applying any conversion of atomic operations and add a customized conversion to our preprocessing script. For this to work, we manually ported the atomic functions from CUDA to DPC++ in a specific header file which is properly added during the postprocessing step.

\subsection{Workaround for Code Similarity}
\label{sec:cuda_familiar}
\gko was originally designed as a GPU-centric sparse linear algebra library using the CUDA programming language and CUDA design patterns for implementing GPU kernels. The \gko HIP backend for targeting AMD GPUs was deployed for production in early 2020. The next step is to support Intel GPUs via a DPC++ backend. Thus, for historic reasons and simplified maintenance, we prefer to keep the coding style of the initial version of the DPC++ backend of \gko similar to the CUDA coding style. We acknowledge that this design choice may narrow down the tasking power of the SYCL language, but consider this design choice as acceptable since task-based algorithms are currently outside the focus of the \gko library at the backend level. However, the \gko library design allows to move closer to the SYCL programming style at a later point if the algorithm properties favor this.
For now, we aim for a high level of code similarity by not only adding the customized cooperative group interface as discussed in \Cref{sec:dpct_workaround}, but also adding a dim3 implementation layer for DPC++ kernel launches that uses the same parameters and parameter order like CUDA and HIP. The interface layer simply reverses the launch parameter order in a library-private member function.

\begin{figure}[!h]
    \centering
    \includegraphics[width=0.7\textwidth]{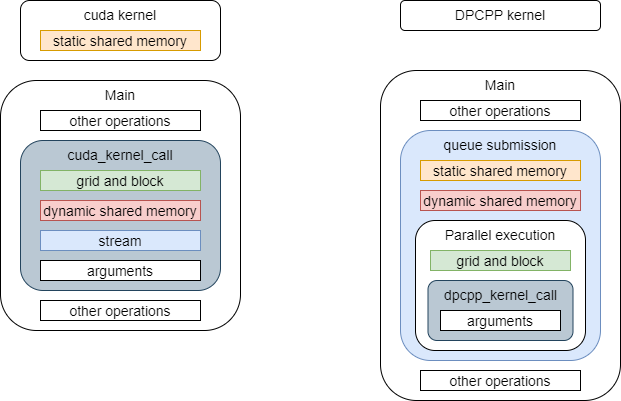}
    \caption{Hierarchical view of usual CUDA (left) and DPC++ (right) kernel call and parameters.}
    \label{fig:dpct_conversion}
\end{figure}

\begin{figure}[!h]
    \centering
    \includegraphics[width=0.7\textwidth]{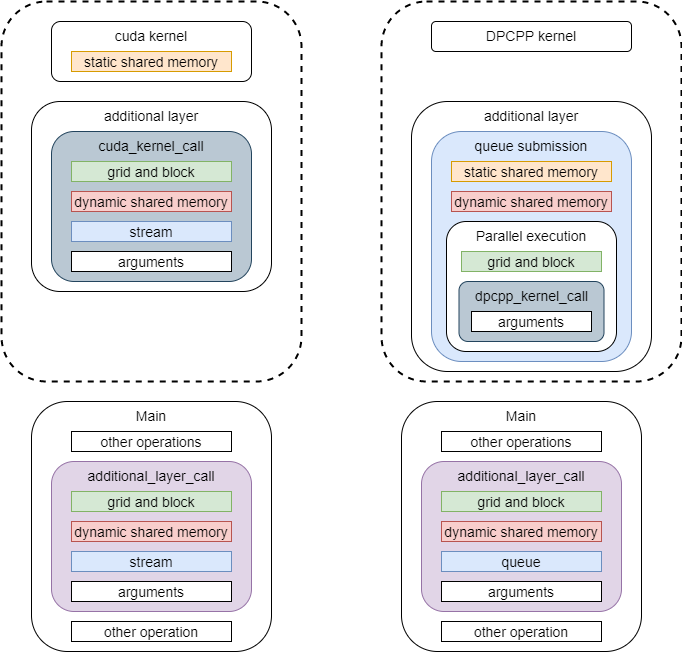}
    \caption{Wrapping the hardware-specific kernels written in HIP, CUDA, and DPC++ into an intermediate layer enables consistency in the kernel invocation across all backends.}
    \label{fig:dpct_conversion_second}
\end{figure}

Despite adding a dim3 helper to use the grid and block notation from CUDA, several differences are left when calling CUDA and DPC++ kernels as in \Cref{fig:dpct_conversion}. One fundamental difference between the CUDA/ HIP ecosystem and DPC++ is that the latter handle the static/dynamic memory allocation in the main component. CUDA and HIP handle the allocation of static shared memory inside the kernel and the allocation of dynamic shared memory in the kernel launch parameters. Another issue is that widely different syntax are used to call CUDA and DPC++ kernels, since DPC++ relies on a hierarchy of calls first to a queue, then a parallel instantiation. For consistency, we add another layer that abstracts the combination of DPC++ memory allocation and DPC++ kernel invocation away from the user. This enables a similar interface for CUDA, HIP, and DPC++ kernels for the main component, and shared memory allocations can be perceived as a kernel feature, see \Cref{fig:dpct_conversion_second}. The purple block (additional\_layer\_call) in \Cref{fig:dpct_conversion_second} has the same structure as the gray block (cuda\_kernel\_call) in the left side of \Cref{fig:dpct_conversion}. Our script will convert the code from the left side of \Cref{fig:dpct_conversion} to the right side of \Cref{fig:dpct_conversion_second} by adding the corresponding additional layer automatically.

\section{Central Sparse Linear Algebra Functionality}
\label{sec:numerics}

An important routine in sparse linear algebra is the \textbf{sparse matrix product} (\spmv). This kernel reflects how a discretized linear operator acts on a vector, and therewith plays the central role in the iterative solution of linear problems and eigenvalue problems. Popular methods based on the repetitive application of the \spmv kernel are Krylov subspace solver such as Conjugate Gradient (CG), GMRES, or BiCGSTAB~\cite{Saad:1986:GGM:14063.14074}, and the PageRank algorithm based on the Power Iteration~\cite{pagerank}. The \spmv kernel is also a key routine in graph analytics as it can be used to identify all immediate neighbors of a node or a set of nodes. 

The sparse data format used to store the discretized matrix and the kernel processing scheme of an \spmv kernel are usually optimized to the hardware characteristics and the matrix properties. In particular on SIMD-parallel architectures like GPUs, the optimization balances between minimization of the matrix memory footprint and efficient parallel processing~\cite{spmv}. 
In the performance evaluation in this paper, we consider two sparse matrix formats: 1) the ``coordinate format'' (\coo) that stores all nonzero entries (and only those) of the matrix along with their column-indices and row-indices, and the ``compressed sparse row'' (\csr) format that reduces the memory footprint of the \coo format further by replacing the explicit row-indices with pointers to the first element in each row of a row-sorted \coo matrix. We focus on these popular matrix formats not only because of their widespread use, but also because Intel's oneMKL library provides a heavily-optimized \csr-\spmv routine for Intel GPUs. For a theoretical analysis of the arithmetic intensity of the sparse data formats, one usually simplifies the \csr memory footprint as 1 floating point value + 1 index value per nonzero entry (8 Byte for single precision \csr, 12 Byte for double precision \csr) and the \coo memory footprint as 1 floating point value + 2 index values per nonzero entry (12 Byte for single precision \csr, 16 Byte for double precision \csr).

Aside from the \spmv kernel which forms the backbone of many algorithms, in the present performance evaluation we also consider \textbf{iterative sparse linear system solvers} that are popular in scientific computing. Specifically, we consider the Krylov solvers CG, BiCGSTAB, CGS, an GMRES. All these solvers are based on the principle of successively building up a Krylov search space and approximating the solution in the Krylov subspace. While the generation of the Krylov search directions is specific to the distinct solvers and realized via a combination of orthogonalizations and vector updates, all solvers heavily rely on the \spmv kernel. All solvers except the GMRES solver are based on short recurrences, that is, the new Krylov search direction is only orthogonalized against the previous search direction~\cite{saad}. Conversely, GMRES stores all search directions, and each new search direction is orthogonalized against all previous search direction~\cite{Saad:1986:GGM:14063.14074}. Therefore, the orthogonalization plays a more important role in the GMRES algorithm. Another difference is that all algorithms except the CG algorithm are designed to solve general linear problems, while the CG algorithm is designed to solve symmetric positive definite problems. 
For a more comprehensive background on the Krylov solvers we consider, we refer the reader to~\cite{saad}.

\section{Experimental Performance Assessment}
\label{sec:experiments}

\subsection{Experiment Setup}
In this paper, we consider two Intel GPUs: the generation 9 (\gnine{}) integrated GPU UHD Graphics P630 with a theoretical bandwidth of 41.6~GB/s
and the generation 12 Intel\textsuperscript{\tiny\textregistered} Iris\textsuperscript{\tiny\textregistered} Xe Max discrete GPU (\gtwelve{})\footnote{\url{https://ark.intel.com/content/www/us/en/ark/products/211013/intel-iris-xe-max-graphics-96-eu.html}} which features 96 execution units and a theoretical bandwidth of 68~GB/s.
To better assess the performance of either GPUs, we include in our analysis the performance we can achieve in bandwidth tests, performance tests, and sparse linear algebra kernels. We note that the \gtwelve architecture lacks native support for IEEE 754 double precision arithmetic, and can only emulate double precision arithmetic. Obviously, emulating double precision arithmetic provides significantly lower performance. Given that native support for double precision arithmetic is expected for future Intel GPUs and using the double precision emulation would artificially degrade the performance results while not providing insight whether \gkos algorithms are suitable for Intel GPUs, we use single precision arithmetic in the performance evaluation on the \gtwelve architecture\footnote{\gko is designed to compile for IEEE 754 double precision arithmetic, single precision arithmetic, double precision complex arithmetic, and single precision complex arithmetic.}.
The DPC++ version we use in all experiments is Intel oneAPI DPC++ Compiler 2021.1 (2020.10.0.1113).  All experiments were conducted on hardware that is part of the Intel DevCloud.

\begin{figure}[!h]
    \centering
 \begin{tabular}{lr}
  \includegraphics[width=.45\columnwidth]{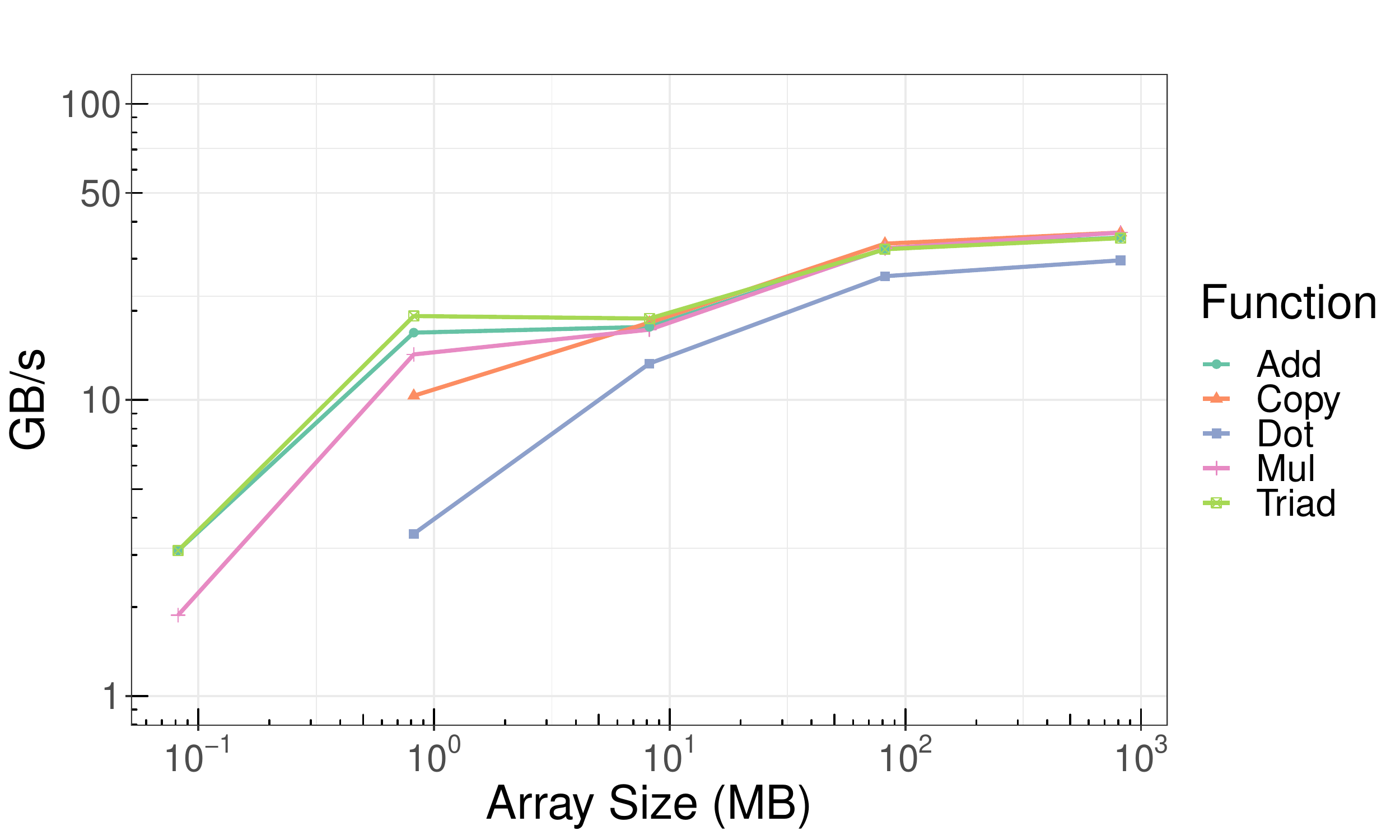}
  &
  \includegraphics[width=.45\columnwidth]{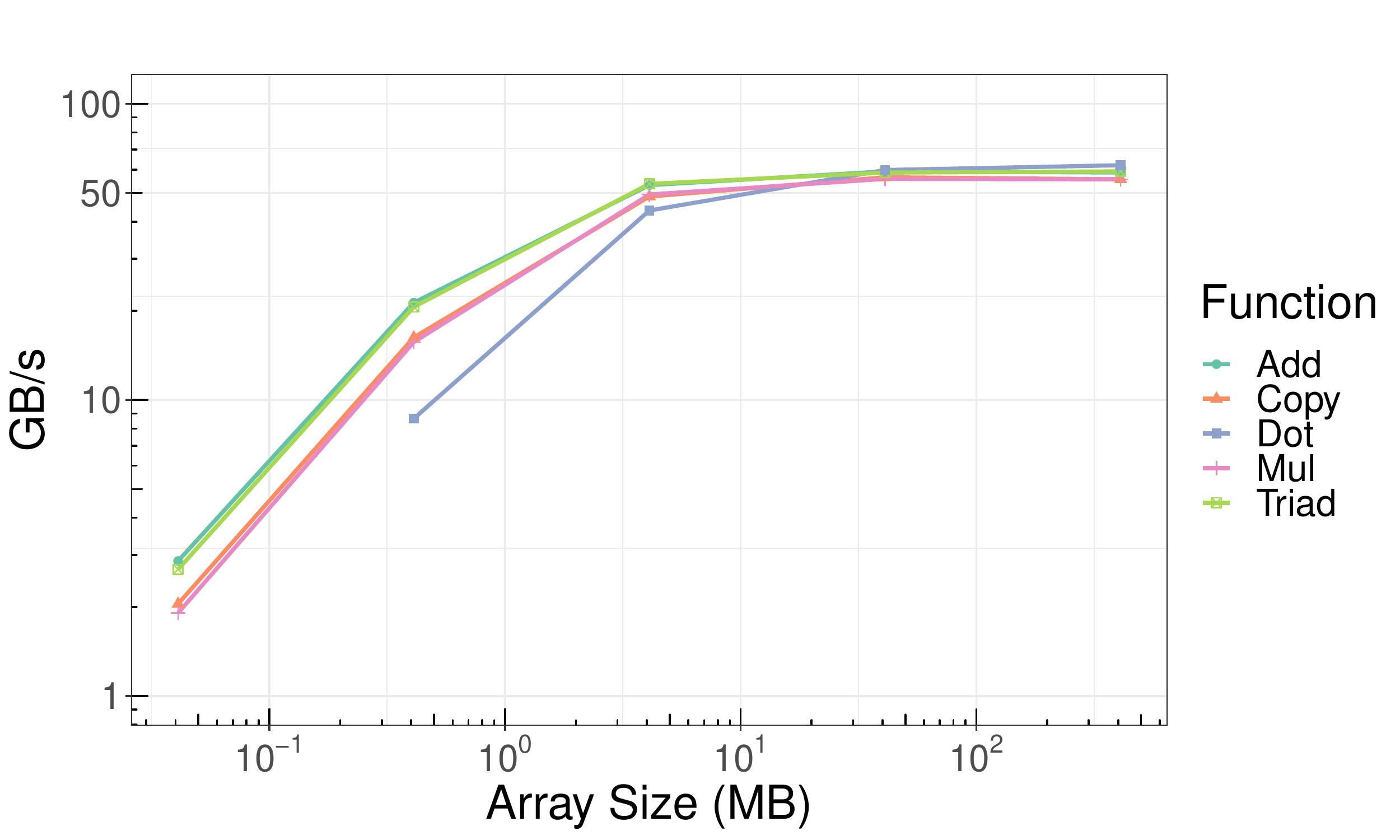}
  \end{tabular}
    \caption{Performance evaluation of the Intel GPUs using the BabelStream benchmark. The bandwidth analysis on the \gnine architecture (left) uses IEEE 754 double precision values, the bandwidth analysis on the \gtwelve architecture (right) uses IEEE 754 single precision values.}
    \label{fig:babelstream}
\end{figure}

\begin{figure}[!h]
    \centering
     \begin{tabular}{lr}
  \includegraphics[width=.45\columnwidth]{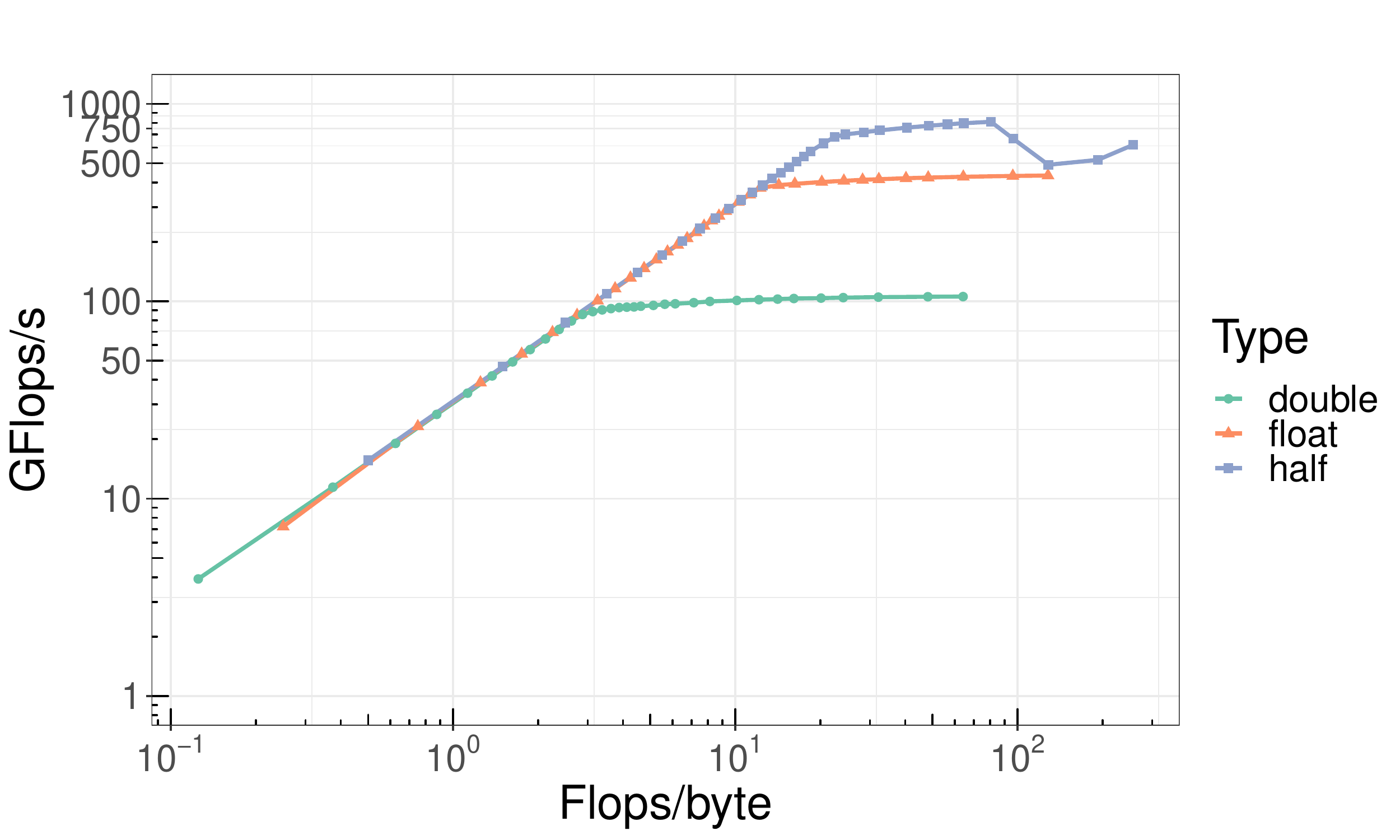}
  &
  \includegraphics[width=.45\columnwidth]{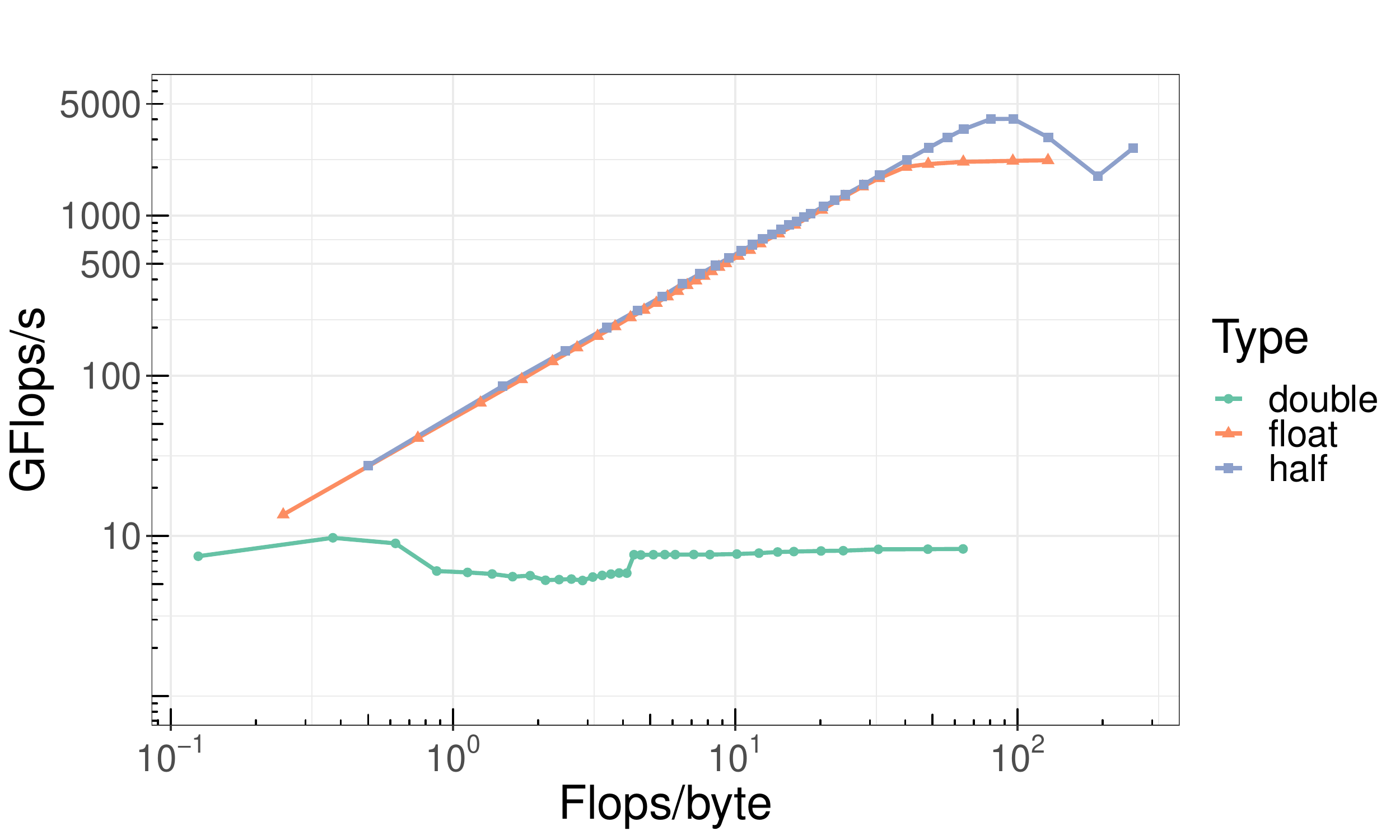}
  \end{tabular}
    \caption{Experimental performance roofline of the Intel GPUs using the mixbench benchmark for the \gnine (left) and \gtwelve (right) GPUs.}
    \label{fig:roofline}
\end{figure}

\subsection{Bandwidth Tests and Roofline Performance Model}
Initially, we evaluate the two GPUs in terms of architecture-specific performance bounds. For that purpose, we use the BabelStream~\cite{babelstream} benchmark to evaluate the peak bandwidth, and the mixbench~\cite{mixbench} benchmark to evaluate the arithmetic performance in different precision formats and derive a roofline model~\cite{roofline}.
In \Cref{fig:babelstream} we visualize the bandwidth we achieve for different memory-intense operations. On both architectures, the {\sc Dot} kernel requiring a global synchronization achieves lower bandwidth than the other kernels. We furthermore note that the \gtwelve architecture achieves for large array sizes about 58 GB/s, which is about 1.6$\times$ the \gnine bandwidth (37 GB/s).

In \Cref{fig:roofline} we visualize the experimental performance roofline for the two GPU architectures. The \gnine architecture achieves about 105 GFLOP/s, 430 GFLOP/s, and 810 GFLOP/s for IEEE double precision, single precision, and half precision arithmetic, respectively. The \gtwelve architecture does not provide native support for IEEE double precision and the double precision emulation achieves only 8 GFLOP/s, which is significantly below the \gnine performance. On the other hand, the \gtwelve architecture achieves 2.2 TFLOP/s and 4.0 TFLOP/s for single precision and half precision floating point operations.

\subsection{\spmv Performance Analysis}

\begin{figure}[!h]
    \centering
    \subfloat[\gnine\label{fig:spmvnine}
    ]{\includegraphics[width=.6\columnwidth]{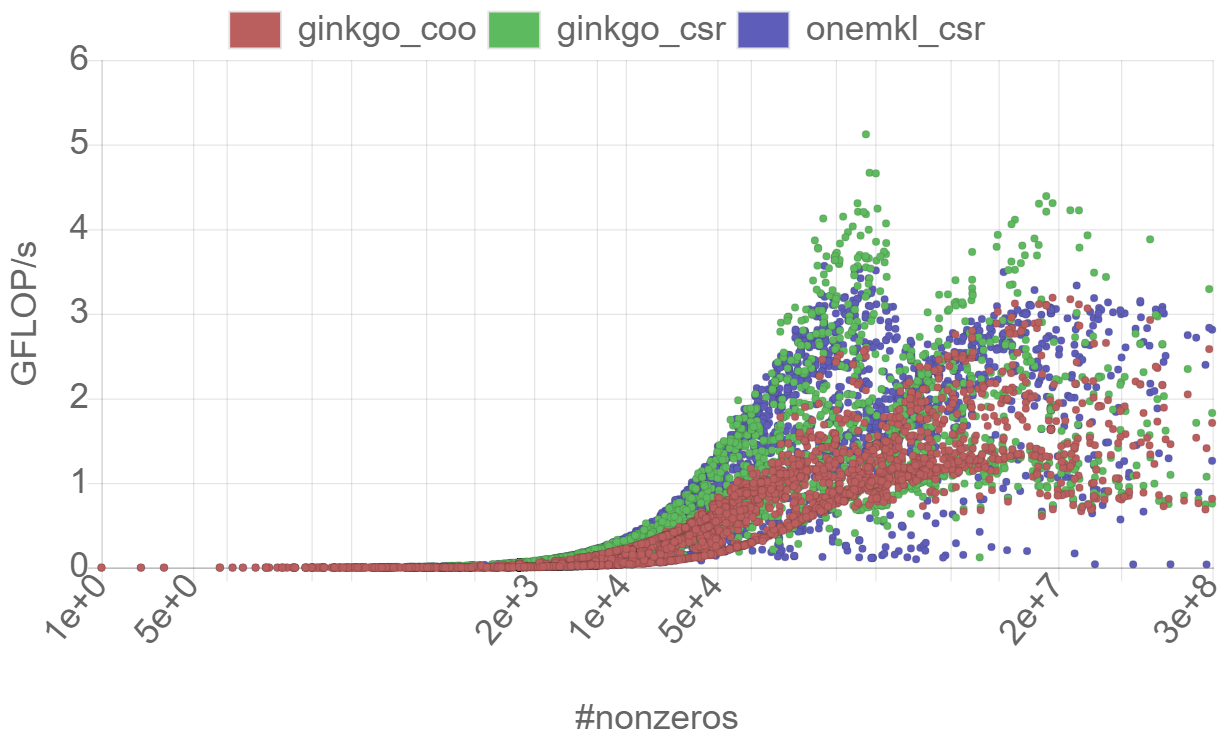}
    }\\
    \subfloat[\gtwelve\label{fig:spmvtwelve}
    ]{\includegraphics[width=.6\columnwidth]{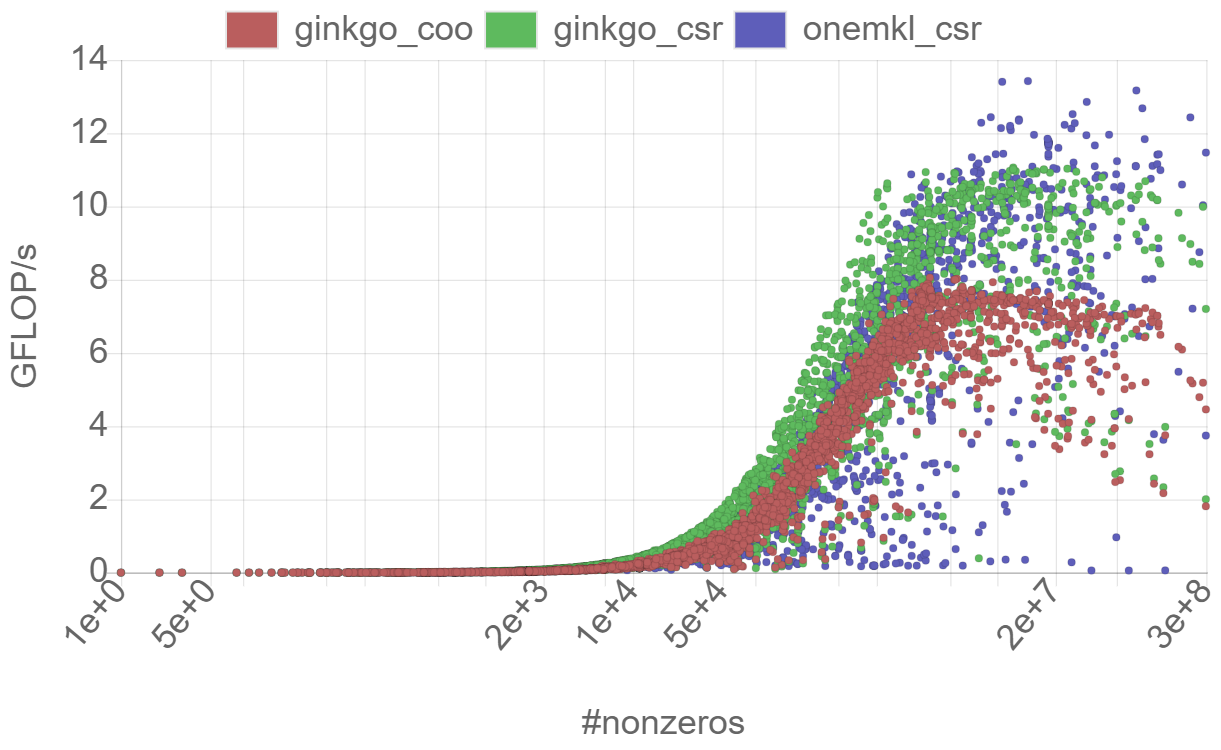}
    }
    \caption{Performance evaluation of two \spmv kernels available in the \gko open source library and Intel's oneMKL vendor library on the Intel GPUs. The experiments on the \gnine architecture (left) use IEEE 754 double precision arithmetic, the experiments on the \gtwelve (right) use IEEE 754 single precision arithmetic.}
    \label{fig:spmvperf}
\end{figure}

Next, we turn to evaluating the performance of numerical functionality on the Intel GPUs. All \spmv experimental performance data we report reflects the average of 10 kernel repetitions after 2 warmup kernel launches. In \Cref{fig:spmvperf}, we visualize the performance of the \csr and \coo \spmv kernels of the \gko library along with the performance of the \csr \spmv kernel from the oneAPI library. Each dot represents the performance for one of the test matrices of the Suite Sparse Matrix Collection~\cite{suitesparse}. On the \gnine GPU, we run these benchmarks using IEEE 754 double precision arithmetic. \gkos \csr \spmv kernel and the \csr \spmv kernel of Intel's oneMKL library achieve similar performance, while \gkos \coo \spmv generally achieves lower performance. Assuming an arithmetic intensity of 1/6 (2 FLOP / 12 Byte) for the double precision \csr \spmv and 1/8 (2 FLOP / 16 Byte) for the double precision \coo \spmv, we can derive for the \gnine architecture (experimental peak bandwidth 37 GB/s) an upper bound for \spmv performance of 6 and 4.6 GFLOP/s respectively. This theoretical upper bound does neither account for the row-pointer overhead in the \csr format nor for the read and write access to the vector. Hence, the experimental performance achieving 5.1 GFLOP/s (\csr) and 3.8 GFLOP/s (\coo) indicate the high efficiency of the \spmv kernel implementations.

Given the lack of native IEEE 754 double precision support, we use IEEE 754 single precision in the performance evaluation on the \gtwelve architecture. Ignoring the access to the vectors and the CSR row-pointer, the arithmetic intensity of the \spmv routines becomes 1/4 (2 FLOP / 8 Byte) for the single precision \csr \spmv and 1/6 (2 FLOP / 12 Byte) for the single precision \coo \spmv. With the experimental bandwidth peak of 58 GB/s, we derive the theoretical performance limits of 14.5 GFLOP/s and 9.7 GFLOP/s for the single precision \csr and \coo \spmv kernels, respectively. The experimental data presented in \Cref{fig:spmvtwelve} reveals that both the \csr and \coo \spmv routines from \gko and the \csr \spmv kernel shipping with Intel's oneAPI library achieve performance close to this theoretical performance limit\footnote{At the point of writing, oneMKL \csr can only operate on shared memory on the \gtwelve architecture and oneMKL does not support the \coo \spmv operation yet.}.

\begin{table}[]
    \centering
    \begin{tabular}{|l|l|r|r|}
    \hline
    \hline
    Matrix & Origin & Size (n) & Nonzeros (nz)\\
    \hline
     rajat31    &  Circuit Simulation Problem & 4,690,002 & 20,316,253\\
     atmosmodj    &  	CFD Problem & 1,270,432 & 8,814,880\\
     nlpkkt160 & Nonlinear Programming Problem & 8,345,600 & 225,422,112\\
     thermal2 & Unstructured FEM & 1,228,045 & 8,580,313\\
     CurlCurl\_4 & 2nd order Maxwell & 2,380,515 & 26,515,867\\
     Bump\_2911 & 3D Geomechanical Simulation & 2,911,419 & 127,729,899\\
     Cube\_Coup\_dt0 & 3D Consolidation Problem & 2,164,760 & 124,406,070\\
     StocF-1456 & Flow in Porous Medium & 1,465,137 &  	21,005,389 \\
     circuit5M &   	Circuit Simulation Problem & 	5,558,326 &  	59,524,291 \\
     FullChip &  Circuit Simulation Problem & 	2,987,012 &  	26,621,990\\
     \hline
     \hline
    \end{tabular}
    \caption{Test matrix along with key characteristics.}
    \label{tab:matrices}
\end{table}

\begin{figure}[!h]
    \centering
    \subfloat[\gnine\label{fig:solvernine}
    ]{\includegraphics[width=.6\columnwidth]{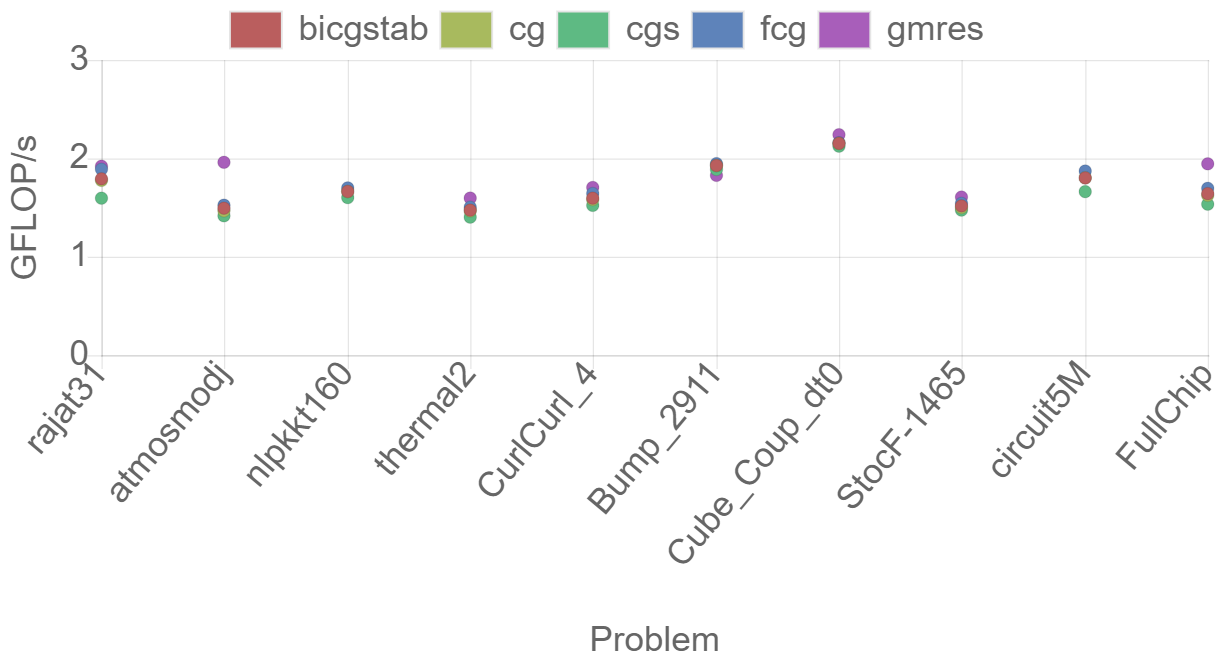}
    }\\
    \subfloat[\gtwelve\label{fig:solvertwelve}
    ]{\includegraphics[width=.6\columnwidth]{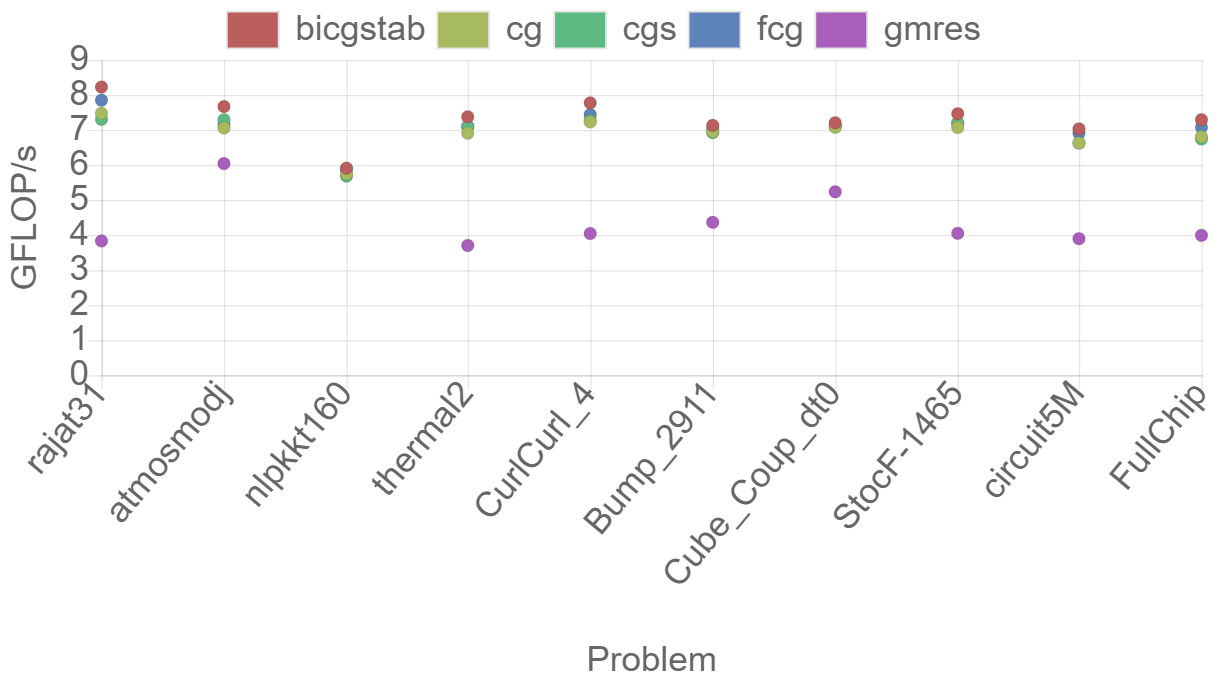}
    }
    \caption{Performance evaluation of \gkos Krylov solvers on the Intel GPUs.}
    \label{fig:solverperf}
\end{figure}
\subsection{Krylov Solver Performance Analysis}
We now turn to complete linear solver applications as they are typical for scientific simulation codes. We run the solver experiment for 1,000 solver iterations after a warm-up phase. The iterative Krylov solvers we consider all have the \spmv kernel as central building block, and we use \gkos \coo \spmv kernel in the solver performance assessment. For this experiment, we select a set of test matrices from the Suite Sparse Matrix Collection that are orthogonal in their characteristics and origin, see \Cref{tab:matrices}. The upper graph in \Cref{fig:solverperf} visualizes the performance for the Krylov solvers on the \gnine architecture. All solvers achieve between 1.5 GFLOP/s and 2.5 GFLOP/s depending on the test matrix. We notice that the performance differences in-between the solvers are quite small compared the performance differences for the distinct problems. The lower graph in \Cref{fig:solverperf} visualizes the performance for the Krylov solvers on the \gtwelve architecture. We recall that \gtwelve does not provide native support for IEEE double precision computations, and we therefore run the solver benchmarks in IEEE single precision.
Overall, in this experiment, the \gko solvers achieve between 5 GFLOP/s and 9 GFLOP/s for the distinct systems. We note that all Krylov solvers based on short recurrences are very similar in terms of performance, while the performance of the GMRES solver is usually significantly lower. This may be due to the fact that the GMRES algorithm requires solving the Hessenberg system, and some needed functionality not yet being supported on the \gtwelve architecture by oneAPI. The developed workaround occurs to achieve lower performance.

\subsection{Platform Portability}
\begin{figure}[!h]
    \centering
    \subfloat[RadeonVII
    ]{\includegraphics[width=.48\columnwidth]{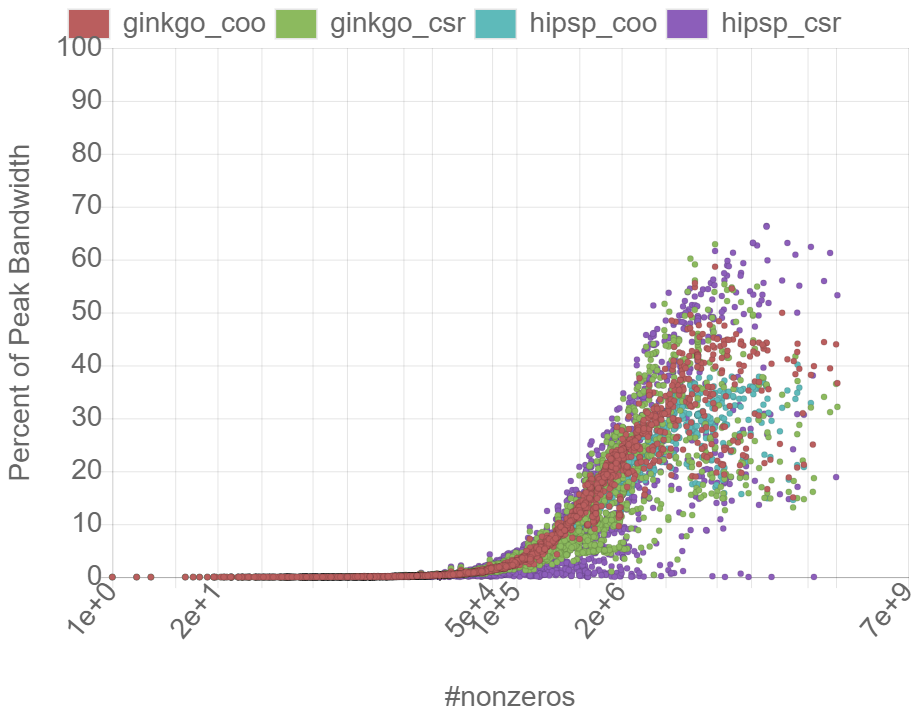}
    }
    \subfloat[V100 
    ]{\includegraphics[width=.48\columnwidth]{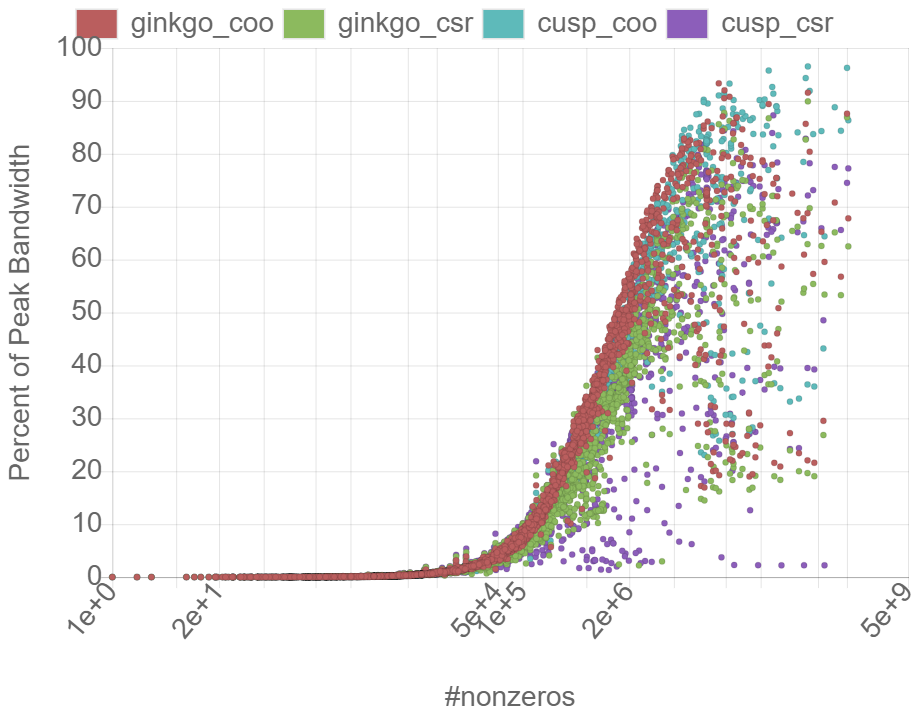}
    }\\
    \subfloat[\gnine\label{fig:spmvnine_percent}
    ]{\includegraphics[width=.48\columnwidth]{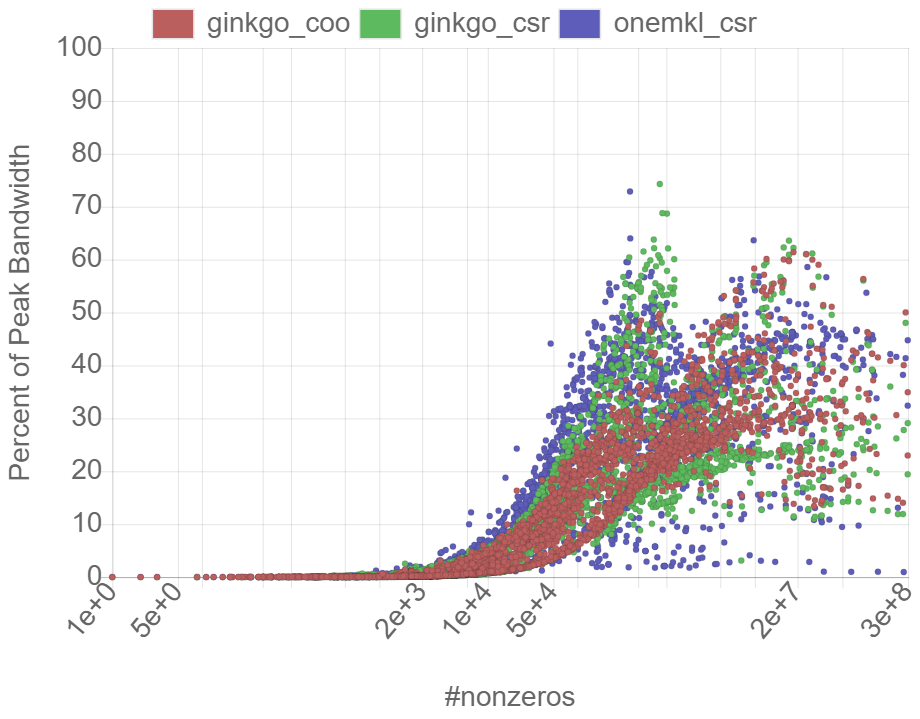}
    }
    \subfloat[\gtwelve\label{fig:spmvtwelve_percent}
    ]{\includegraphics[width=.48\columnwidth]{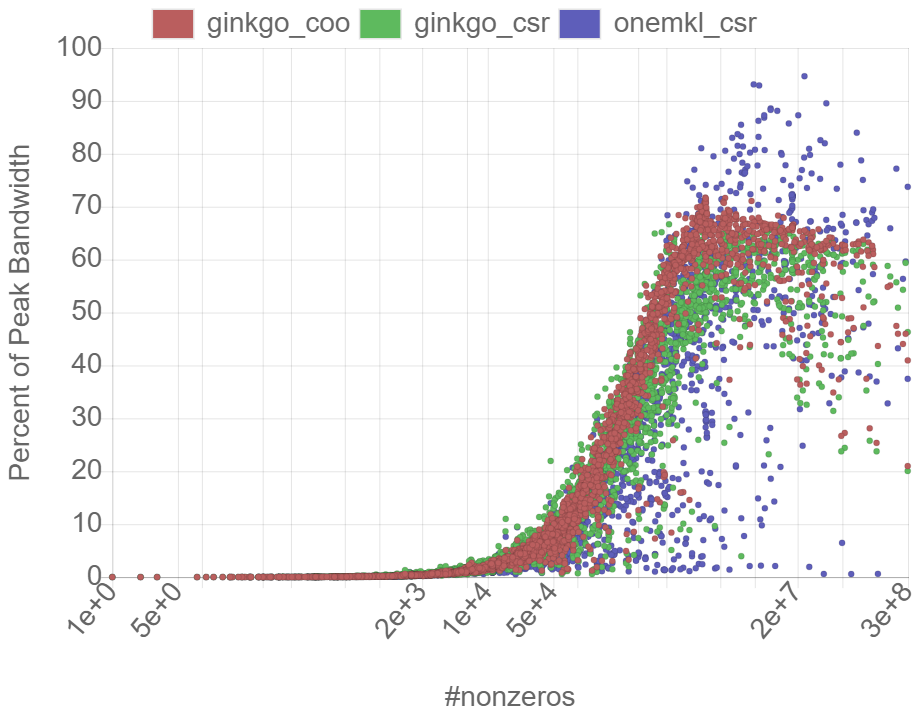}
    }

    \caption{\spmv kernel bandwidth relative to the peak bandwidth for \spmv kernels available in the \gko open source library and vendor libraries on the AMD, NVIDIA, and Intel GPUs.}
    \label{fig:spmvpercent}
\end{figure}

Finally, we want to take a look at the platform portability of \gko{}'s functionality, and see whether the ``dpcpp'' backend can provide the same efficiency like the ``cuda'' and ``hip'' backends. For that, we do not focus on the absolute performance the functionality achieves on GPUs from AMD, NVIDIA, and Intel, but the relative performance taking the theoretical performance limits reported in the GPU specifications as baseline. This approach reflects the aspect that the GPUs differ significantly in their performance characteristics, and that Intel's OneAPI ecosystem and Intel's high performance GPU architectures still being under active development and not yet having reached the maturity level of other GPU computing ecosystems. At the same time, reporting the performance relative to the theoretical limits allows to quantify the suitability of \gko{}'s algorithms and efficiency of \gko{}'s kernel implementations for the distinct GPU architectures. It may also indicate the performance we can expect for \gko{}'s functionality when scaling up the GPU performance.
In \Cref{fig:spmvpercent} we report the relative performance of different \spmv kernels on the AMD Radeon VII (``hip'' backend), the NVIDIA V100 (``cuda'' backend), and the Intel \gnine and Intel \gtwelve GPUs (both ``dpcpp'' backend).

As expected, the achieved bandwidth heavily depends on the \spmv kernel and the characteristics of the test matrix. Overall, the performance figures indicate that the \spmv kernels achieve about 90\% of peak bandwidth on A100 and \gtwelve, but about 60-70\% of peak bandwidth on RadeonVII and \gnine. At the same time, we notice that on the \gtwelve, the performance of the oneMKL \csr \spmv to be inconsistent, largely outperforming \gko{}'s \spmv kernels for some cases, but underperforming for others. Overall, \gko{}'s \spmv kernels are on all platforms competitive to the vendor libraries, indicating the validity of the library design and demonstrating good performance portability.

\section{Summary and Outlook}
\label{sec:conclusions}
In this paper, we have presented an open source math library featuring a DPC++ backend to execute on Intel GPUs. We elaborated on the porting effort and the workarounds we implemented to enable DPC++ support. We also evaluated the raw performance of different Intel GPU generations and investigated how this raw performance translates into the developed basic sparse linear algebra operations and sparse iterative solvers. The performance analysis revealed that DPC++ allows to achieve high efficiency in terms of translating raw performance into mathematical algorithms. The portability analysis shows \gkos performance portability on modern HPC platforms. Future work will focus on running the platform-portable DPC++ kernels on AMD GPUs and NVIDIA GPUs and compare the kernel performance with the performance of kernels written in the vendor-specific programming languages HIP and CUDA, respectively. We failed to include the work in this paper as at the time of writing, platform portability of DPC++ is not yet enabled.


\FloatBarrier
\bibliographystyle{plain}
\bibliography{references}

\begin{thebibliography}{10}

\bibitem{gpe}
Hartwig Anzt, Yen-Chen Chen, Terry Cojean, Jack Dongarra, Goran Flegar, Pratik
  Nayak, Enrique~S. Quintana-Ort\'{\i}, Yuhsiang~M. Tsai, and Weichung Wang.
\newblock Towards continuous benchmarking: An automated performance evaluation
  framework for high performance software.
\newblock In {\em Proceedings of the Platform for Advanced Scientific Computing
  Conference}, PASC '19, pages 9:1--9:11, New York, NY, USA, 2019. ACM.

\bibitem{ginkgojoss}
Hartwig Anzt, Terry Cojean, Yen-Chen Chen, Goran Flegar, Fritz Göbel, Thomas
  Grützmacher, Pratik Nayak, Tobias Ribizel, and Yu-Hsiang Tsai.
\newblock Ginkgo: A high performance numerical linear algebra library.
\newblock {\em Journal of Open Source Software}, 5(52):2260, 2020.

\bibitem{ginkgoarxiv}
Hartwig Anzt, Terry Cojean, Goran Flegar, Fritz Goebel, Thomas Gruetzmacher,
  Pratik Nayak, Tobias Ribizel, Yu-Hsiang Tsai, and Enrique~S Quintana-Orti.
\newblock Ginkgo: A modern linear operator algebra framework for high
  performance computing.
\newblock {\em arXiv preprint arXiv:2006.16852}, 2020.

\bibitem{cojean2020ginkgo}
Terry Cojean, Yu-Hsiang~"Mike" Tsai, and Hartwig Anzt.
\newblock Ginkgo -- a math library designed for platform portability, 2020.

\bibitem{babelstream}
T.~Deakin, J.~Price, Matt Martineau, and Simon McIntosh-Smith.
\newblock Evaluating attainable memory bandwidth of parallel programming models
  via babelstream.
\newblock {\em International Journal of Computational Science and Engineering},
  17:247--262, 2017.

\bibitem{toms-adaptiveblockjacobi}
Goran Flegar, Hartwig Anzt, Terry Cojean, and Enrique~S Quintana-Ort{\'\i}.
\newblock Adaptive precision block-jacobi for high performance preconditioning
  in the ginkgo linear algebra software.
\newblock {\em ACM Transaction on Mathematical Software}, 47(2), 2021.

\bibitem{googletest}
Google Inc.
\newblock Googletest.
\newblock \url{https://github.com/google/googletest}, 2021.

\bibitem{sycl}
Ronan Keryell, Ruyman Reyes, and Lee Howes.
\newblock Khronos sycl for opencl: A tutorial.
\newblock In {\em Proceedings of the 3rd International Workshop on OpenCL},
  IWOCL '15, New York, NY, USA, 2015. Association for Computing Machinery.

\bibitem{mixbench}
Elias Konstantinidis and Yiannis Cotronis.
\newblock A quantitative roofline model for gpu kernel performance estimation
  using micro-benchmarks and hardware metric profiling.
\newblock {\em Journal of Parallel and Distributed Computing}, 107:37 -- 56,
  2017.

\bibitem{pagerank}
L.~Page, S.~Brin, R.~Motwani, and T.~Winograd.
\newblock The pagerank citation ranking: Bringing order to the web.
\newblock In {\em Proceedings of the 7th International World Wide Web
  Conference}, pages 161--172, Brisbane, Australia, 1998.

\bibitem{Saad:1986:GGM:14063.14074}
Youcef Saad and Martin~H Schultz.
\newblock {GMRES: a generalized minimal residual algorithm for solving
  nonsymmetric linear systems}.
\newblock {\em SIAM J. Sci. Stat. Comput.}, 7:856--869, July 1986.

\bibitem{saad}
Youssef. Saad.
\newblock {\em Iterative Methods for Sparse Linear Systems}.
\newblock SIAM, 2nd edition, 2003.

\bibitem{suitesparse}
SuiteSparse.
\newblock {Matrix Collection}.
\newblock \url{https://sparse.tamu.edu}, 2018.
\newblock {Accessed in April 2018}.

\bibitem{Tsai2020PreparingGF}
Yuhsiang~M. Tsai, Terry Cojean, Tobias Ribizel, and Hartwig Anzt.
\newblock "preparing ginkgo for amd gpus -- a testimonial on porting cuda code
  to hip".
\newblock In Bartosz Balis, Dora B.~Heras, Laura Antonelli, Andrea Bracciali,
  Thomas Gruber, Jin Hyun-Wook, Michael Kuhn, Stephen~L. Scott, Didem Unat, and
  Roman Wyrzykowski, editors, {\em "Euro-Par 2020: Parallel Processing
  Workshops"}, pages 109--121, Cham, 2021. "Springer International Publishing".

\bibitem{spmv}
Sam Williams, Nathan Bell, Jee Choi, Michael Garland, Leonid Oliker, and
  Richard Vuduc.
\newblock Sparse matrix vector multiplication on multicore and accelerator
  systems.
\newblock In Jakub Kurzak, David~A. Bader, and Jack Dongarra, editors, {\em
  Scientific Computing with Multicore Processors and Accelerators}. CRC Press,
  2010.

\bibitem{roofline}
Samuel Williams, Andrew Waterman, and David Patterson.
\newblock {Roofline: An Insightful Visual Performance Model for Multicore
  Architectures}.
\newblock {\em Commun. ACM}, 52(4):65--76, April 2009.

\end{thebibliography}

\end{document}